\documentclass[%
 reprint,
 superscriptaddress,
 showpacs,preprintnumbers,
 amsmath,amssymb,
 aps,
 prl,
 longbibliography,
lengthcheck,%
]{revtex4-1}

\usepackage[dvipdfmx]{graphicx}
\usepackage{dcolumn}
\usepackage{bm}
\usepackage{color}
\usepackage{ulem}
\usepackage{hyperref}

\begin{document}

\preprint{APS/123-QED}

\title{
Peltier effect of phonon driven by ac electromagnetic waves
}

\author{Hiroaki Ishizuka}
\affiliation{
Department of Physics, Tokyo Institute of Technology, Meguro, Tokyo, 152-8551, Japan
}

\author{Masahiro Sato}
\affiliation{
Department of Physics, Chiba University, Chiba 263-8522, Japan
}

\date{\today}

\begin{abstract}
Steady current in metals induces a thermal gradient, a phenomenon known as the Peltier effect.
The Peltier effect is one of the fundamental phenomena in the thermoelectric properties of materials and is also used in applications such as refrigerators. 
In this work, we show that an analogous phenomenon occurs by phonons in a material subject to linearly-polarized light.
Under light illumination, an energy current of phonons occurs through a nonlinear optical effect similar to the bulk photovoltaic effect.
We formulate the nonlinear Peltier coefficient of the photogalvanic energy current carried by phonons using nonlinear response theory.
From the general formula, we show that the photogalvanic energy current occurs only in a non-centrosymmetric system with two or more optical phonon bands.
We demonstrate the generation of the photogalvanic energy current using a one-dimensional ion chain with three ions in a unit cell, which predicts the generation of an experimentally observable energy current using available THz-infrared light sources. 
\end{abstract}

\pacs{
}

\maketitle

\begin{figure}
  \includegraphics[width=\linewidth]{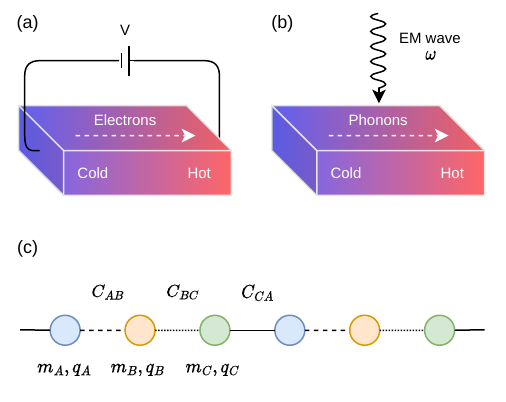}
  \caption{
  Schematic of (a) Peltier effect in $n$-type semiconductor and (b) that by phonon current in insulators.
  A temperature gradient is induced by electric current in the Peltier effect, whereas phonon current induces the temperature gradient in the Peltier effect of phonons.
  (c) An example of the non-centrosymmetric insulator in which the Peltier effect of phonon occurs. The blue, orange, and green balls represent $A$, $B$, and $C$ sublattice ions, respectively.
  The $m_a$ and $q_a$ ($a=A,B,C$) are, respectively, the mass and charge of $a$th ion, and $C_{ab}$ ($a,b=A,B,C$) are the strength of quadratic coupling between $a$ and $b$ ions.
  }\label{fig:model}
\end{figure}

{\it Introduction} ---
In semiconductors, a steady electric current causes a temperature gradient in a material, which is known as the Peltier effect [Fig.~\ref{fig:model}(a)]~\cite{Goupil2016a,Abrikosov2017a}.
The thermoelectric properties of materials are not only important from the viewpoint of transport phenomena but also for applications such as refrigerators.
As the thermoelectric effects are related to the energy current carried by the electric carriers, analogous phenomena can occur by a flow of other quasi-particles, such as magnons and phonons.
In fact, a spin current analog of the Seebeck effect called the spin Seebeck effect is observed in magnetic insulators~\cite{Uchida2010a,Hirobe2017a}, in which case magnons and spinons carry the spin angular momentum.
On the other hand, as accelerating magnons and phonons by the electromagnetic field is difficult, generating magnon and phonon current is often done by introducing thermal gradient, such as in the thermal Hall effect experiment~\cite{Strohm2005a,Sheng2006a,Kagan2008a,Fujimoto2009a,Katsura2010a,Onose2010a,Matsumoto2011a,Mori2014a,Hirschberger2015a,Ideue2017a,Kasahara2018a,Grissonnanche2019a,Saito2019a}.
Therefore, an analog of the Peltier effect, that is, externally controlling the temperature gradient by inducing the flux of quasiparticles, remains challenging.

A possible solution to controlling the flux of magnons and phonons is to utilize non-linear optical phenomena in non-centrosymmetric materials.
The nonlinear response of bulk materials, especially the bulk photovoltaic effect, has received renewed attention from the viewpoint of application and the non-trivial contribution of electronic structures, such as the Berry phase~\cite{Kraut1979a,Sturman1992a,Sipe2000a,Young2012a,Cook2017a,Tokura2018a}. 
In addition, recent studies on the optical response in magnetic materials found that 
the spin current of magnons~\cite{Proskurin2018a,Ishizuka2019b,Ishizuka2022a} and spinons~\cite{Ishizuka2019a} can be induced by a nonlinear response.
As these carriers also carry energy, the Peltier effect of charge-neutral particles may also occur by light illumination [Fig.~\ref{fig:model}(b)].
Among the charge-neutral quasiparticles in materials, phonons are promising in this prospect as the heat transport in materials is often dominated by phonons~\cite{Abrikosov2017a,Sologubenko2001,Kawamata2008}.
The larger contribution to thermal properties promises a relatively larger non-linear effect. 
Hence, a nonlinear response of phonons might be a route to realizing novel thermal functionalities. 

In this work, we explore the possibility of the Peltier effect of phonons and attempt to understand its basic properties.
Using the nonlinear response theory for bosons~\cite{Ishizuka2022a} as a reference, 
we formulate a general theory for the light-induced energy current.
Based on this theory, we argue that a dc phonon current occurs by illuminating a linearly polarized light to noncentrosymmetric insulators.
Unlike the bulk photovoltaic and magnon photovoltaic effects, we show that at least two optical modes (more than three phonon modes, including acoustic modes) are necessary for realizing the Peltier effect of phonons.
In the last, using a minimal one-dimensional model, we argue that the magnitude of energy current induced by this mechanism is comparable to those observed in the heat conductivity measurement.

{\it Nonlinear Peltier coefficient} --- 
The temperature gradient occurs if a flow of energy or heat occurs in a material.
The flow is described by energy current density, which is defined by the continuum equation $\partial_t \rho_E(\bm x,t)+\nabla\cdot\bm J_Q(\bm x,t)=0$.
Here, $\rho_E(\bm x,t)$ and $\bm J_Q(\bm x,t)$ are the energy and energy current densities at position $\bm x$ and time $t$, respectively.
Hence, evaluating the Peltier effect reduces to evaluating the average energy current flowing in the material.
Phenomenologically, the energy current induced by a nonlinear optical effect reads
\begin{align}
    J_Q^\lambda(\Omega)=\sum_{\mu,\nu}\Pi^{(2)}_{\lambda;\mu\nu}(\Omega;\omega,\Omega-\omega)E_\mu(\omega)E_\nu(\Omega-\omega).\label{eq:Peltier}
\end{align}
Here $J_Q^\lambda(\Omega)=\int J_Q^\lambda(t)e^{-{\rm i}\Omega t}dt$ and $E_\mu(\omega)=\int E_\mu(t)e^{-{\rm i}\omega t}dt$ are respectively the Fourier transform of the spatially averaged energy current density $J_Q^\lambda(t)$ 
and the $\mu$ component of the spatially 
uniform electric field $E_\mu(t)$ in applied electromagnetic waves. 
The averaged current density is defined as  
$J_Q^\lambda(t)=\frac1V\int J_Q^\lambda(\bm x,t) dx^d$  with $V$ being the system volume and $d$ being the dimension of the system. 
The frequency of induced energy current is $\Omega$, and the frequency of the incident electromagnetic wave is $\omega$.
Equation~\eqref{eq:Peltier} defines the nonlinear Peltier coefficient $\Pi^{(2)}_{\lambda;\mu\nu}$, which we study in the rest of this paper.

We note that the nonlinear Peltier effect occurs only in non-centrosymmetric phonon systems.
This is shown by the symmetry argument.
By acting the inversion operation, the energy current and electric field transforms as $J_Q\to-J_Q$ and $E_\nu\to-E_\nu$.
Hence, Eq.~\eqref{eq:Peltier} transforms $J_Q^\lambda=\Pi^{(2)}_{\lambda;\mu\nu}E_\mu E_\nu \to J_Q^\lambda=-\Pi^{(2)}_{\lambda;\mu\nu}E_\mu E_\nu$.
Therefore, similar to the photovoltaic effect in semiconductors, $\Pi^{(2)}_{\lambda;\mu\nu}=0$ in centrosymmetric systems.

To study the basic properties of energy current carried by phonons, we consider a general low-energy Hamiltonian for phonons of a crystal in $d$ dimension with $n_{uc}$ atoms in a unit cell,
\begin{align}
H=\sum_{ia\mu}\frac{\hat p^2_{ia\mu}}{2m_a}+\sum_{ia\mu,jb\nu}\hat u_{ia\mu}A_{ia\mu,jb\nu}\hat u_{jb\nu}.\label{eq:H}
\end{align}
Here, $A_{ia\mu,jb\nu}$ is the coupling constant, $u_{ia\mu}$ is the displacement along the $\mu$ axis of $a\,(=1,\cdots,n_{uc})$ sublattice atom in $i$th unit cell, and $p_{ia\mu}$ is the conjugate momentum of $u_{ia\mu}$ satisfying $[u_{ia\mu},p_{jb\nu}]={\rm i}\hbar\delta_{ij}\delta_{ab}\delta_{\mu\nu}$; $\hbar$ is the Dirac constant.
The excitation of $H$ is described by free bosons called phonon~\cite{Grosso2013a}.
Using the phonon representation, $H$ reads
\begin{align}
H=\sum_{n,\bm k}\hbar\omega_{n\bm k}(\hat b_{n\bm k}^\dagger\hat b_{n\bm k}+\frac12),
\end{align}
where $\hat b_{n\bm k}$ ($\hat b_{n\bm k}^\dagger$) is the annihilation (creation) operator of a phonon with band index $n$ and momentum $\bm k$, and $\omega_{n\bm k}$ is the phonon frequency.
Here, we define $n$ such that $\omega_{n\bm k}\le \omega_{m\bm k}$ when $n<m$.

We investigate the Peltier effect in this phonon system by calculating the energy current induced by an ac field. 
For the sake of generality, we consider an ac perturbation.
\begin{align}
&H'=\sum_\mu B^\mu E_\mu(t),\\
&\hat B^\mu=\sum_{n,\bm k}\beta^\mu_{n\bm k}\hat b_{n\bm k}+(\beta^\mu_{n\bm k})^\ast\hat b_{n\bm k}^\dagger.\label{eq:H'}
\end{align}
This form of perturbation includes most of the basic coupling between the ions and the electromagnetic field.
For instance, the coupling of ion charge to the {\it uniform} electric field $H'=-\sum_{i,a,\eta}q_aE_\eta(t) \hat u_{ia\eta}$ reads
\begin{align}
\beta_{n\bm 0}=&\sum_{a,\mu}q_aE_\mu(t)|n\bm 0\rangle_{a\mu} \sqrt{\frac{\hbar N}{2m_a\omega_{n\bm 0}}},\label{eq:coupling}
\end{align}
and $\beta_{n\bm k\ne\bm0}=0$. Here, $N$ is the size of the system, $|n\bm k\rangle$ is the $n$th eigenmode of dynamical matrix
$\tilde A_{a\mu,b\nu}(\bm k)$, which is an $n_{uc}d \times n_{uc}d$ matrix whose elements are $\tilde A_{a\mu,b\nu}(\bm k)=\sum_i\frac{A_{ia\mu,0b\nu}}{\sqrt{m_am_b}}e^{-{\rm i}\bm k\cdot(\bm r_{ia}-\bm r_{0b})}$; it corresponds to the eigenvector of $n$th phonon mode, i.e., $\omega^2_{n\bm k}|n\bm k\rangle=\tilde A_{a\mu,b\nu}(\bm k)|n\bm k\rangle$.

A general formula for $\bm J_Q$ carried by phonons is given in Ref.~\cite{Hardy1963a}, in which the formula is quadratic in the phonon creation and annihilation operators.
For the phonon Hamiltonian of Eq.~\eqref{eq:H}, the $\lambda$ component of the energy current operator reads~\cite{Hardy1963a}
\begin{align}
\hat J_Q^\lambda&=\sum_{n,m,\bm k}\hat b_{n\bm k}^\dagger v_{nm}^\lambda(\bm k)\hat b_{m\bm k}+\sum_{n,m,\bm k}\hat b_{n\bm k}^\dagger v_{n\bar m}^\lambda(\bm k)\hat b_{m-\bm k}^\dagger\nonumber\\
&+\sum_{n,m,\bm k}\hat b_{n-\bm k} v_{\bar nm}^\lambda(\bm k)\hat b_{m\bm k}+\sum_{n,m,\bm k}\hat b_{n-\bm k} v_{\bar n\bar m}^\lambda(\bm k)\hat b_{m-\bm k}^\dagger.\label{eq:J}
\end{align}
where
\begin{align}
  v^\lambda_{nm}(\bm k)=&\frac{\hbar(\omega_{n\bm k}+\omega_{m\bm k})\langle n\bm k|\partial_{k_\lambda}\tilde A(\bm k)|m\bm k\rangle}{8V\sqrt{\omega_{n\bm k}\omega_{m\bm k}}},\\
  v^\lambda_{n\bar m}(\bm k)=&\frac{\hbar(\omega_{n\bm k}-\omega_{m\bm k})\langle n\bm k|\partial_{k_\lambda}\tilde A(\bm k)|m\bm k\rangle}{8V\sqrt{\omega_{n\bm k}\omega_{m\bm k}}},\\
  v^\lambda_{\bar nm}(\bm k)=&-\frac{\hbar(\omega_{n\bm k}-\omega_{m\bm k})\langle n\bm k|\partial_{k_\lambda}\tilde A(\bm k)|m\bm k\rangle}{8V\sqrt{\omega_{n\bm k}\omega_{m\bm k}}},\\
  v^\lambda_{\bar n\bar m}(\bm k)=&-\frac{\hbar(\omega_{n\bm k}+\omega_{m\bm k})\langle n\bm k|\partial_{k_\lambda}\tilde A(\bm k)|m\bm k\rangle}{8V\sqrt{\omega_{n\bm k}\omega_{m\bm k}}}.
\end{align}
We note that, in the above equation, we can always take $v^\lambda_{nm}(\bm k)=v^\lambda_{\bar m\bar n}(-\bm k)$, $v^\lambda_{n\bar m}(\bm k)=v^\lambda_{m\bar n}(-\bm k)$, and $v^\lambda_{\bar nm}(\bm k)=v^\lambda_{\bar mn}(-\bm k)$ without reducing the generality. 
For the sake of convenience, we call $v^\lambda_{nn}(\bm k)$ and $v^\lambda_{\bar n\bar n}(\bm k)$ the intra-band elements of velocity matrix, and the other terms the inter-band elements. 

The formula for the nonlinear Peltier coefficient is obtained by extending the nonlinear-response theory~\cite{Ishizuka2022a,Suppl}.
The formula for the dc ($\Omega=0$) Peltier coefficient at temperature $T$ reads
\begin{align}
&\Pi^{(2)}_{\lambda;\mu\nu}(0;\omega,-\omega)=\nonumber\\
  &-\frac1{2\pi\hbar^2}\sum_{n,m}\frac{1}{\omega-\omega_{n\bm0}-\frac{\rm i}{2\tau}}\frac{\beta_{n\bm0}^\mu[v_{n\bar m}(\bm0)+v_{m\bar n}(\bm0)]\beta_{m\bm0}^\nu}{\omega_{n\bm0}+\omega_{m\bm0}+\frac{\rm i}{2\tau}}\nonumber\\
  &+\frac1{2\pi\hbar^2}\sum_{n,m}\frac{1}{\omega+\omega_{n\bm0}-\frac{\rm i}{2\tau}}\frac{(\beta_{n\bm0}^\mu)^\ast[v_{\bar nm}(\bm0)+v_{\bar mn}(\bm0)](\beta_{m\bm0}^\nu)^\ast}{\omega_{n\bm0}+\omega_{m\bm0}-\frac{\rm i}{2\tau}}\nonumber\\
  &+\frac1{2\pi\hbar^2}\sum_{n,m}\frac{1}{\omega-\omega_{n\bm0}-\frac{\rm i}{2\tau}}\frac{\beta_{n\bm0}^\mu[v_{nm}(\bm0)+v_{\bar m\bar n}(\bm0)](\beta_{m\bm0}^\nu)^\ast}{\omega_{n\bm0}-\omega_{m\bm0}+\frac{\rm i}{2\tau}}\nonumber\\
  &-\frac1{2\pi\hbar^2}\sum_{n,m}\frac{1}{\omega+\omega_{n\bm0}-\frac{\rm i}{2\tau}}\frac{(\beta_{n\bm0}^\mu)^\ast[v_{mn}(\bm0)+v_{\bar n\bar m}(\bm0)]\beta_{m\bm0}^\nu}{\omega_{n\bm0}-\omega_{m\bm0}-\frac{\rm i}{2\tau}}.\label{eq:Pi}
\end{align}
Here, $\tau=\tau(T)$ is the phenomenological phonon lifetime at $T$.
The general formula with nonzero $\Omega\ne0$ is also given in the Supplemental Material~\cite{Suppl}.

\begin{figure}
  \includegraphics[width=\linewidth]{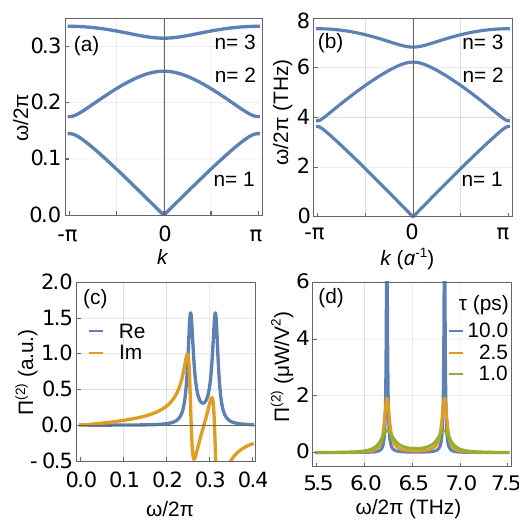}
  \caption{
  Phonon bands of the three-ion model with (a) $C_{AB}=C_{BC}=C_{CA}=1$, $M_A=2/3$, $M_B=1$, $M_C=4/3$, and $a=1$, and (b) $C_{AB}=C_{CA}=50$ kg/s$^2$, $C_{BC}=40$ kg/s$^2$, $M_A=48$ Da, $M_B=50$ Da, $M_C=52$ Da, and $a=4$ \AA.
  (c,d) The $\omega$ dependence of nonlinear Peltier coefficient $\Pi^{(2)}_{x;xx}(0;\omega,-\omega)$. (c) The real and imaginary parts of $\Pi^{(2)}_{x;xx}(0;\omega,-\omega)$ for the model in (a) with the relaxation time $\tau=10$, and (d) the relaxation-time dependence of $\Pi^{(2)}_{x;xx}(0;\omega,-\omega)$ for the model in (d).
  }\label{fig:Pi}
\end{figure}

{\it Three-ion chain} ---
As a demonstration, we consider a one-dimensional lattice model with three ions in a unit cell whose ions move only along the chain direction (Fig.~\ref{fig:model}), i.e., we assume that there are only longitudinal modes.
The Hamiltonian reads,
\begin{align}
H=\sum_{i,a}\frac{\hat p_{ia}^2}{2M_a}+\frac12\sum_{\langle ia,jb\rangle}C_{ab}(\hat u_{ia}-\hat u_{jb})^2,
\end{align}
where $\hat u_{ia}$ is the displacement of an atom at $a=A,B,C$ sublattice of $i$th unit cell from its equilibrium position, $\hat p_{ia}$ is the momentum conjugate to $\hat u_{ia}$, $M_a$ is the mass of atom at $a$ sublattice, and $C_{ab}$ is the coupling constant between the nearest-neighbor $a$ and $b$ sublattice atoms [Fig.~\ref{fig:model}(c)]. 
This model has one acoustic and two optical modes, as shown in Fig.~\ref{fig:Pi}(a) and (b). 

Using this model, we computed the phonon Peltier effect induced by the coupling of ion charges to the electric field of ac electromagnetic waves given in Eq.~\eqref{eq:H'}.
Figure~\ref{fig:Pi}(c) and \ref{fig:Pi}(d) show the $\omega$ dependence of nonlinear Peltier coefficient for dc energy current, $\Pi^{(2)}_{1;11}(0,\omega,-\omega)$.
The Peltier effect of phonons occurs when the frequency of incident light matches the energy of an optical mode ($\omega=\omega_{2\bm0},\omega_{3\bm0}$ for the cases in Fig.~\ref{fig:Pi}), similar to the resonance effect. However, no peak exists at $\omega=0$, where it corresponds to the energy of the acoustic mode.

{\it Absence of coupling to the acoustic modes} ---
To understand the absence of a resonance peak at $\omega=0$, we look into the acoustic mode terms, $n=1,\cdots$. 
In the case of a translationally-symmetric system, $\beta^\mu_{n\bm0}$ for acoustic modes become $\beta^\mu_{n\bm0}\propto\sum_b q_b$ for the $H'$ in Eq.~\eqref{eq:H'}~\cite{Suppl}.
Hence, in Eq.~\eqref{eq:Pi}, we can effectively neglect the contribution from acoustic modes in a charge-neutral system, i.e., when $\sum_a q_a=0$.
In this case, the Peltier coefficient reads
\begin{align}
&\Pi^{(2)}_{\lambda;\mu\nu}(0;\omega,-\omega)=\nonumber\\
  &-\frac1{2\pi\hbar^2}\sum_{n,m}^{opt}\frac{1}{\omega-\omega_{n\bm0}-\frac{\rm i}{2\tau}}\frac{\beta_{n\bm0}^\mu[v_{n\bar m}(\bm0)+v_{m\bar n}(\bm0)]\beta_{m\bm0}^\nu}{\omega_{n\bm0}+\omega_{m\bm0}+\frac{\rm i}{2\tau}}\nonumber\\
  &+\frac1{2\pi\hbar^2}\sum_{n,m}^{opt}\frac{1}{\omega+\omega_{n\bm0}-\frac{\rm i}{2\tau}}\frac{(\beta_{n\bm0}^\mu)^\ast[v_{\bar nm}(\bm0)+v_{\bar mn}(\bm0)](\beta_{m\bm0}^\nu)^\ast}{\omega_{n\bm0}+\omega_{m\bm0}-\frac{\rm i}{2\tau}}\nonumber\\
  &+\frac1{2\pi\hbar^2}\sum_{n,m}^{opt}\frac{1}{\omega-\omega_{n\bm0}-\frac{\rm i}{2\tau}}\frac{\beta_{n\bm0}^\mu[v_{nm}(\bm0)+v_{\bar m\bar n}(\bm0)](\beta_{m\bm0}^\nu)^\ast}{\omega_{n\bm0}-\omega_{m\bm0}+\frac{\rm i}{2\tau}}\nonumber\\
  &-\frac1{2\pi\hbar^2}\sum_{n,m}^{opt}\frac{1}{\omega+\omega_{n\bm0}-\frac{\rm i}{2\tau}}\frac{(\beta_{n\bm0}^\mu)^\ast[v_{mn}(\bm0)+v_{\bar n\bar m}(\bm0)]\beta_{m\bm0}^\nu}{\omega_{n\bm0}-\omega_{m\bm0}-\frac{\rm i}{2\tau}},\label{eq:Pi2}
\end{align}
where the sum is over the optical phonon bands.
Thereby, the nonlinear Peltier effect occurs only in materials with optical modes.

{\it Inter-band terms in velocity} ---
We also note that the phonon dispersion in Fig.~\ref{fig:Pi}(a) is symmetric about the $k=0$ line, which indicates that the group velocity of excited phonons, $\partial_k\omega_{nk}$, is zero at $k=0$; hence, the diagonal terms in Eq.~\eqref{eq:J} is zero.
The observation implies that the nonlinear Peltier effect occurs not by the selective excitation of phonons with a finite velocity but by some other mechanism.
In fact, in the study of the bulk photovoltaic effect, a photocurrent proportional to the inter-band elements of the velocity matrix is known, such as the shift current~\cite{Kraut1979a,Sturman1992a}.
A similar contribution to the photogalvanic spin current is also known~\cite{Ishizuka2019a,Ishizuka2019b}.
A slight difference, however, exists: the inter-band elements related to two optical modes appear in the phonon Peltier effect, as inferred from the numerators of Eq.~\eqref{eq:Pi2}.
In contrast, the inter-band elements of the valence and conduction bands involved in the optical transition appear in the case of shift current.
The phonon Peltier effect discussed here is another phenomenon related to the non-trivial optical transition, not described by the selective excitation of a phonon with a finite velocity.

As the phonon energy current is related to the inter-band elements, multiple optical modes are necessary for realizing the Peltier effect of phonons (the acoustic modes do not contribute to the Peltier effect as discussed above).
In fact, for the Hamiltonian in Eq.~\eqref{eq:H}, one can show that the group velocity at $\bm k=0$ is zero if the phonon bands are non-degenerate at $\bm k=0$.
From a physical viewpoint, the vanishing diagonal terms manifest time-reversal symmetry.
Hence, two or more optical bands [at least three bands, including the acoustic mode(s)] are necessary for realizing the Peltier effect of phonons.

{\it Relaxation-time dependence} --- To gain further insight into the nature of the nonlinear Peltier effect, we next look into the relaxation-time dependence which is relevant to the temperature dependence and magnitude of the Peltier effect. In Eq.~\eqref{eq:Pi}, the temperature dependence appears in the relaxation time $\tau$. Hence, understanding the temperature dependence of $\Pi^{(2)}$ reduces to analyzing the $\tau$ dependence of $\Pi^{(2)}$. Figures~\ref{fig:Pi}(b) and \ref{fig:Pi}(d) shows the $\Pi^{(2)}$ with different relaxation-time $\tau$. The figures show a monotonic increase of $\Pi^{(2)}$ with increasing $\tau$ when $\omega=\omega_{n0}$. Indeed, at $\omega=\pm\omega_{n0}$ and $\omega_{nk}\tau\gg1$, the real part of $\Pi^{(2)}$ in Eq.~\eqref{eq:Pi} reads
\begin{align}
&\text{Re}[\Pi^{(2)}_{\lambda;\mu\nu}(0;\pm\omega_n,\mp\omega_n)]\sim\nonumber\\
  &\frac{2\tau}{\pi\hbar^2}\sum_{m(\ne n)}\frac{\text{Im}[\beta_{n\bm0}^\mu v_{n\bar m}^\lambda(\bm0)\beta_{m\bm0}^\nu]}{\omega_{n\bm0}+\omega_{m\bm0}}
  -\frac{\text{Im}[\beta_{n\bm0}^\mu v_{nm}^\lambda(\bm0)(\beta_{m\bm0}^\nu)^\ast]}{\omega_{n\bm0}-\omega_{m\bm0}}.
\end{align}
The $\tau$-linear dependence is distinct from that of shift current, which is independent of $\tau$~\cite{Kraut1979a,Sipe2000a}.
Rather, it resembles the photogalvanic spin current using magneto-resonance effect~\cite{Ishizuka2022a}, whose spin current conductivity is proportional to $\tau$.
Experimentally investigating the $\tau$ dependence via the temperature dependence of $\tau$ may provide a route to experimentally delineating the nonlinear Peltier effect from other phenomena.

In the $\tau\to\infty$ limit, the Peltier coefficient reads
\begin{align}
  &\Pi_{\lambda;\mu\nu}^{(2)}(0;\omega,-\omega)=\sum_{m\ne n}^{opt}\frac{\sqrt{\omega_{\bm 0m}\omega_{\bm 0n}}}{2V\hbar}\beta_{m\bm0}^\mu a_{mn}^\lambda(\bm0) \beta_{n\bm0}^\nu\times\nonumber\\
  &\hspace{16mm}\left[-\delta(\omega+\omega_{m\bm0})-\delta(\omega-\omega_{m\bm0})\right.\nonumber\\
  &\hspace{16mm}\qquad\left.+{\rm i}{\cal P}\frac1{\omega+\omega_{m\bm0}}+{\rm i}{\cal P}\frac1{\omega-\omega_{m\bm0}}\right],
\end{align}
where $a_{nm}^\lambda(\bm k)={\rm i}\langle n\bm k|\partial_{k_\lambda}|m\bm k\rangle$ is the non-abelian Berry connection of phonons, defined in a similar manner to the abelian Berry connection~\cite{Zhang2010,Qin2011}.
Here, we used the fact that $a_{nm}^\lambda(\bm 0)$ and $\beta_{b\bm0}^\nu$ are real, which holds for Eqs.~\eqref{eq:H} and \eqref{eq:coupling}.
The real part of $\Pi_{\lambda;\mu\nu}^{(2)}(0;\omega,-\omega)$ shows a sharp peak at $\omega=\omega_{m\bm0}$, as expected from the relaxation-time dependence.

{\it Magnitude of nonlinear Peltier effect} --- In the last, we discuss the magnitude of the energy current and the possibility of experimental observation.
The results in Fig.~\ref{fig:Pi}(d) indicate that the Peltier coefficient is around $\Pi^{(2)}\sim 1 \mu$W/V$^2$ at the resonance frequency, which is typically in THz to infrared range.
Therefore, assuming the relative electrical permittivity $\epsilon=10$, the energy current density induced by an ac electric field of $|E|=10^5$ V/m is $J_E\sim 10^{-2}$ W/cm$^2$.
This result should be compared to the energy current measured in thermal conductivity experiments.
The energy current density induced by the temperature gradient can be estimated from the thermal conductivity $\kappa$.
In the case of an insulator with $\kappa\sim0.1-10$ W/mK, the thermal gradient of $\Delta T=10^3$ K/m induces $J_E\sim 10^{-2}-10^0$ W/cm$^2$.
As the energy current in our estimate is similar to those measured in thermal transport experiments, the Peltier effect of phonons should induce an observable temperature gradient in candidate materials.

{\it Summary} ---
In this work, we theoretically studied the possibility of the Peltier effect of phonons induced by the illumination of THz to infrared electromagnetic waves.
In this phenomenon, the flow of phonons induces energy current, which results in the thermal gradient.
However, unlike the Peltier effect by electric current, the phonon current is driven by a nonlinear response similar to the bulk photovoltaic effect.
To formulate the Peltier effect, we focused on the energy current of phonons induced by the illumination of electromagnetic waves.
The general formula for energy current in the second order of the electromagnetic waves is derived using the nonlinear response theory.
This formula is directly applicable to arbitrary phonon models. 
Using the formula, we generally showed that at least two optical phonon modes are necessary for inducing the energy current, and the coupling to acoustic mode does not contribute to the phonon Peltier effect.
In the last, we demonstrated the Peltier effect using a three-ion model whose phonon bands consist of one acoustic and two optical modes.
The result shows that the Peltier effect of phonons occurs when the frequency of the incident electromagnetic wave matches the frequency of optical modes.

Recently, thermal imaging techniques have enabled spatial resolution of temperature in small devices~\cite{Uchida2018}.
As the nonlinear Peltier effect induces a temperature gradient in an isolated device, it should be observable using the thermal imaging method.

\acknowledgements
This work is supported by JSPS KAKENHI (Grant Numbers JP19K14649, JP23K03275, JP20H01830, and JP20H01849), and by a Grant-in-Aid for Scientific Research on Innovative Areas “Quantum Liquid Crystals” (Grant No. JP19H05825) and “Evolution of Chiral Materials Science using Helical Light Fields” (Grants No. JP22H05131 and No. JP23H04576) 
from JSPS of Japan.

\clearpage
\onecolumngrid

\setcounter{equation}{0}
\setcounter{figure}{0}
\setcounter{table}{0}
\setcounter{page}{1}
\makeatletter

\renewcommand{\thefigure}{S\arabic{figure}}
\renewcommand{\theequation}{S\arabic{equation}}
\renewcommand{\thetable}{S\arabic{table}}
\renewcommand{\vec}[1]{{\bm #1}}

\begin{center}
\textbf{\large
Supplemental Material to \\
``{\it Peltier effect of phonon driven by ac electromagnetic waves}''
}
\end{center}


\section{Nonlinear response theory}
In this section, we consider a nonlinear response theory for many body systems.
To be concrete, we consider a perturbation
\begin{align}
&H'=\sum_\mu \hat B^\mu F_\mu(t),\label{eq:coupling}
\end{align}
where $\hat B^\mu$ is a many-body operator and $F_\mu(t)$ ($\mu=1, \dots$) are time-dependent external fields.
By extending the linear-response theory to the second-order in the electric field, the Fourier transform of the energy current
\begin{align}
J_Q(\Omega)=\int dt\;J_Q(t)e^{-{\rm i}\omega t},
\end{align}
reads
\begin{align}
J_Q(\Omega)=\int \frac{d\omega}{2\pi}\frac{(\rho_n-\rho_m)B_{nm}^\mu}{\omega+E_n-E_m-{\rm i}/2\tau}\left(\frac{B_{ml}^\nu (J_Q)_{ln}}{\Omega+E_n-E_l-{\rm i}/2\tau}-\frac{(J_Q)_{ml} B_{ln}^\nu}{\Omega+E_l-E_m-{\rm i}/2\tau}\right)F_\mu(\omega)F_\nu(\Omega-\omega),
\end{align}
where
\begin{align}
F_\mu(t)=\int\frac{d\omega}{2\pi}F(\omega)e^{{\rm i}\omega t},
\end{align}
 $E_n$ is the internal many-body energy of the many-body state $|n\rangle$, $\rho_n=e^{-\beta E_n}/Z$ is the statistical probability of the systems being the $n$th state, $B^\mu_{nm}=\langle n|\hat B^\mu|m\rangle$, and $\tau$ is the phenomenological relaxation time.
Comparing the above equation to the definition of $\Pi^{(2)}_{\lambda;\mu\nu}(\Omega;\omega,\Omega-\omega)$, the formula for nonlinear Peltier coefficient reads,
\begin{align}
\Pi_{\lambda;\mu\nu}^{(2)}(\Omega;\omega,\Omega-\omega)=\frac1{2\pi}\sum_{n,m,l}\frac{(\rho_n-\rho_m)B_{nm}^\mu}{\omega+E_n-E_m-{\rm i}/2\tau}\left(\frac{B_{ml}^\nu (J_Q)_{ln}}{\Omega+E_n-E_l-{\rm i}/2\tau}-\frac{(J_Q)_{ml} B_{ln}^\nu}{\Omega+E_l-E_m-{\rm i}/2\tau}\right).\label{eq:nonlin:Pi}
\end{align}

We apply the theory to a free boson system whose Hamiltonian $\hat H$, the operator in Eq.~\eqref{eq:coupling}, $\hat B^\mu$ ($\mu=x,y,z$), and the energy current operator $\hat J_Q$ are given by
  \begin{align}
  &\hat H=\sum_{n,\bm k} \hbar\omega_{n\bm k}(b_{n\bm k}^\dagger b_{n\bm k}+\frac12),\qquad\hat B^\mu=\sum_n\beta^\mu_{n\bm0}\hat b_{n\bm0}+(\beta^\mu_{n\bm0})^\ast\hat b_{n\bm0}^\dagger,\nonumber\\
  &\hat J_Q=\sum_{n,m,\bm k}\hat b_{n\bm k}^\dagger v_{nm}(\bm k)\hat b_{m\bm k}+\sum_{n,m,\bm k}\hat b_{n\bm k}^\dagger v_{n\bar m}(\bm k)\hat b_{m-\bm k}^\dagger+\sum_{n,m,\bm k}\hat b_{n-\bm k} v_{\bar nm}(\bm k)\hat b_{m\bm k}+\sum_{n,m,\bm k}\hat b_{n-\bm k} v_{\bar n\bar m}(\bm k)\hat b_{m-\bm k}^\dagger.\label{eq:nonlin:model}
  \end{align}
Note that $J_Q(t)=\langle \hat J_Q\rangle$.
Here, we assume
$$v_{nm}(\bm k)=v_{\bar m\bar n}(-\bm k),\quad v_{n\bar m}(\bm k)=v_{m\bar n}(-\bm k),\quad v_{\bar nm}(\bm k)=v_{\bar mn}(-\bm k),$$
as the assumptions do not reduce generality.
In addition, the hermiticity of observables require
$$[v_{mn}(\bm k)]^\ast=v_{nm}(\bm k),\quad [v_{m\bar n}(\bm k)]^\ast=v_{\bar nm}(\bm k),\quad [v_{\bar m\bar n}(\bm k)]^\ast=v_{\bar n\bar m}(\bm k).$$
For $\bm k=\bm 0$, the above conditions require
$$v_{nm}(\bm 0)=v_{\bar m\bar n}(\bm 0),\quad v_{n\bar m}(\bm 0)=v_{m\bar n}(\bm 0)=[v_{\bar nm}(\bm 0)]^\ast=[v_{\bar mn}(\bm 0)]^\ast,\quad v_{\bar nm}(\bm 0)=v_{\bar mn}(\bm 0).$$

For the model in Eq.~\eqref{eq:nonlin:model}, the nonlinear Peltier coefficient in Eq.~\eqref{eq:nonlin:Pi} reads
  \begin{align}
  &\Pi_{\lambda;\mu\nu}^{(2)}(\Omega;\omega,\Omega-\omega)=\nonumber\\
  &\frac1{2\pi}\sum_{n,m}\frac{1}{\hbar\omega-\hbar\omega_{n\bm0}-{\rm i}\hbar/2\tau}\frac{\beta_{n\bm0}^\mu[v_{n\bar m}(\bm0)+v_{m\bar n}(\bm0)]\beta_{m\bm0}^\nu}{\hbar\Omega-\hbar\omega_{n\bm0}-\hbar\omega_{m\bm0}-{\rm i}\hbar/2\tau}
  +\frac1{2\pi}\sum_{n,m}\frac{1}{\hbar\omega+\hbar\omega_{n\bm0}-{\rm i}\hbar/2\tau}\frac{(\beta_{n\bm0}^\mu)^\ast[v_{\bar nm}(\bm0)+v_{\bar mn}(\bm0)](\beta_{m\bm0}^\nu)^\ast}{\hbar\Omega+\hbar\omega_{n\bm0}+\hbar\omega_{m\bm0}-{\rm i}\hbar/2\tau}\nonumber\\
  &-\frac1{2\pi}\sum_{n,m}\frac{1}{\hbar\omega-\hbar\omega_{n\bm0}-{\rm i}\hbar/2\tau}\frac{\beta_{n\bm0}^\mu[v_{nm}(\bm0)+v_{\bar m\bar n}(\bm0)](\beta_{m\bm0}^\nu)^\ast}{\hbar\Omega-\hbar\omega_{n\bm0}+\hbar\omega_{m\bm0}-{\rm i}\hbar/2\tau}-\frac1{2\pi}\sum_{n,m}\frac{1}{\hbar\omega+\hbar\omega_{n\bm0}-{\rm i}\hbar/2\tau}\frac{(\beta_{n\bm0}^\mu)^\ast[v_{mn}(\bm0)+v_{\bar n\bar m}(\bm0)]\beta_{m\bm0}^\nu}{\hbar\Omega-\hbar\omega_{m\bm0}+\hbar\omega_{n\bm0}-{\rm i}\hbar/2\tau}.
  \end{align}
Using the property of $v_{nm}$, $v_{n\bar m}$, $v_{\bar nm}$, and $v_{\bar n\bar m}$,
  \begin{align}
  \Pi_{\lambda;\mu\nu}^{(2)}&(\Omega;\omega,\Omega-\omega)=\nonumber\\
  &\frac1{\pi}\sum_{a,b}\frac{1}{\omega-\hbar\omega_{a\bm0}-{\rm i}/2\tau}\frac{\beta_{a\bm0}^\mu v_{a\bar b}(\bm0)\beta_{b\bm0}^\nu}{\Omega-\hbar\omega_{a\bm0}-\hbar\omega_{b\bm0}-{\rm i}/2\tau}+\frac1{\pi}\sum_{a,b}\frac{1}{\omega+\hbar\omega_{a\bm0}-{\rm i}/2\tau}\frac{(\beta_{a\bm0}^\mu v_{a\bar b}(\bm0) \beta_{b\bm0}^\nu)^\ast}{\Omega+\hbar\omega_{a\bm0}+\hbar\omega_{b\bm0}-{\rm i}/2\tau}\nonumber\\
  &-\frac1{\pi}\sum_{a,b}\frac{1}{\omega-\hbar\omega_{a\bm0}-{\rm i}/2\tau}\frac{\beta_{a\bm0}^\mu v_{ab}(\bm0)(\beta_{b\bm0}^\nu)^\ast}{\Omega-\hbar\omega_{a\bm0}+\hbar\omega_{b\bm0}-{\rm i}/2\tau}-\frac1{\pi}\sum_{a,b}\frac{1}{\omega+\hbar\omega_{a\bm0}-{\rm i}/2\tau}\frac{[\beta_{a\bm0}^\mu v_{ab}(\bm0)(\beta_{b\bm0}^\nu)^\ast]^\ast}{\Omega-\hbar\omega_{b\bm0}+\hbar\omega_{a\bm0}-{\rm i}/2\tau}.
  \end{align}
When $\Omega=0$, the dc nonlinear Peltier coefficient becomes
  \begin{align}
  \Pi_{\lambda;\mu\nu}^{(2)}&(0;\omega,-\omega)=\nonumber\\
  &-\frac1{\pi}\sum_{a,b}\frac{1}{-\omega-\hbar\omega_{a\bm0}+{\rm i}/2\tau}\left(\frac{\beta_{a\bm0}^\mu v_{a\bar b}(\bm0) \beta_{b\bm0}^\nu}{\hbar\omega_{a\bm0}+\hbar\omega_{b\bm0}+{\rm i}/2\tau}\right)^\ast+\frac1{\pi}\sum_{a,b}\frac{1}{-\omega+\hbar\omega_{a\bm0}+{\rm i}/2\tau}\frac{\beta_{a\bm0}^\mu v_{a\bar b}(\bm0)\beta_{b\bm0}^\nu}{\hbar\omega_{a\bm0}+\hbar\omega_{b\bm0}+{\rm i}/2\tau}\nonumber\\
  &+\frac1{\pi}\sum_{a,b}\frac{1}{-\omega-\hbar\omega_{a\bm0}+{\rm i}/2\tau}\left(\frac{\beta_{a\bm0}^\mu v_{ab}(\bm0)(\beta_{b\bm0}^\nu)^\ast}{\hbar\omega_{a\bm0}-\hbar\omega_{b\bm0}+{\rm i}/2\tau}\right)^\ast-\frac1{\pi}\sum_{a,b}\frac{1}{-\omega+\hbar\omega_{a\bm0}+{\rm i}/2\tau}\frac{\beta_{a\bm0}^\mu v_{ab}(\bm0)(\beta_{b\bm0}^\nu)^\ast}{\hbar\omega_{a\bm0}-\hbar\omega_{b\bm0}+{\rm i}/2\tau}.\label{eq:nonlin:Pidc}
  \end{align}
Note that, this formula implies $\Pi_{\lambda;\mu\nu}^{(2)}(0;\omega,-\omega)=[\Pi_{\lambda;\mu\nu}^{(2)}(0;-\omega,\omega)]^\ast$.
In the main text, we use Eq.~\eqref{eq:nonlin:Pidc} to study the nonlinear Peltier effect.

We note that, sometimes, $\sum_{\bm k}v_{nn}(\bm k)=0$ condition is required when rewriting the current operator in the form of Eq.~\eqref{eq:nonlin:model}.
For the case of the energy current operator, this usually holds due to the fact that $v_{nn}(\bm k)=(\hbar/4)\nabla_k[\omega^2_{n\bm k}]$ where $\nabla_k=(\partial_{kx},\partial_{ky},\partial_{kz})$.
Hence, the above theory applies to the general quadratic phonon model.

\section{On the property of velocity operator}

The phonon energy current operator for a lattice model is given by Eq.~(7), which is related to the dynamical matrix $\tilde A(\bm k)$~\cite{supHardy1963a}.
From the definition of $\tilde A(\bm k)$, following relations hold
\begin{align}
\tilde A_{ab}(\bm k)=[\tilde A_{ba}(\bm k)]^\ast,\qquad\tilde A_{ab}(-\bm k)=\tilde A_{ba}(\bm k).    
\end{align}
The two equations state that the dynamical matrix $\tilde A(\bm k)$, whose $(a,b)$ element is $\tilde A_{ab}(\bm k)$, is a hermitian matrix, and the $(a,b)$ element of $\tilde A(-\bm k)$ is the complex conjugate of the same element in $\tilde A_{ab}(\bm k)$.
Therefore,  $\partial_{k_\mu}\tilde A(\bm k)=-{}^t[\partial_{q_\mu}\tilde A(\bm q)]_{\bm q=-\bm k}$.
From this relation and the definition in Eq.~(7),
\begin{align}
    v^a_{nm}(\bm k)=-v^a_{mn}(-\bm k).
\end{align}
Hence, $v^a_{nn}(\bm 0)=0$ for the optical modes if the phonon bands are non-degenerate at $\bm k=\bm 0$.

\newcommand{\bibitemS}[1]{
\stepcounter{count}
\bibitem[S\thecount]{#1}}
\newcounter{count}
 %

 %

\end{document}